# Cellulose nanocrystals mimicking micron sized fibers to assess the deposition of latex particles on cotton

**Evdokia K. Oikonomou[1], Konstantin Golemanov[2], Pierre-Emmanuel Dufils[3], James Wilson[3], Ritu Ahuja[2], Laurent Heux,[4] and Jean-François Berret[1]‡**

*[1]Université de Paris, CNRS, Matière et systèmes complexes, 75013 Paris, France*
*[2]Solvay Research & Innovation Center Singapore, 1 Biopolis Drive, Amnios, Singapore 138622*
*[3]Solvay Paris, 52 rue de la Haie-Coq, 93308 Aubervilliers Cedex, France*
*[4]Centre de recherches sur les macromolécules végétales (CERMAV), BP 53, 38041 Grenoble, France*

**Abstract** We report the interactions of cationic latex particles synthesized by RAFT/MADIX-mediated emulsion polymerization with anionic cellulose nanocrystals (CNCs) and cotton fabrics. Latexes in the size range 200-300 nm with poly(butyl acrylate) or poly(2-ethylhexyl acrylate) hydrophobic cores and hydrophilic shell are synthesized. We show that the latex/CNC interaction is mediated by electrostatics, the interaction being the strongest with the most charged particles. The adsorption process is efficient and does not require any functionalization step for either cellulose or latex. A major result is the observation by cryogenic transmission electron microscopy of latexes coated with entangled arrays of CNCs, and for the softer particles a notable deformation of their structure into faceted polyhedra. By labeling the latexes with hydrophobic carbocyanine dyes, their deposition on woven cotton fabrics is studied *in situ* and quantified by fluorescence microscopy. As with the CNCs, the highest deposition on cotton in the wet and dried states is achieved with the most charged latexes. This demonstrates that CNCs can serve as models to adjust the interactions of latex particles with cotton, and thus optimize manufacturing processes for the development of advanced textiles.

‡ Corresponding author: jean-francois.berret@u-paris.fr
Website: https://www.jean-francois-berret-website-pro.fr

## 1. INTRODUCTION

The use of materials derived from renewable resources is currently being considered with the aim of promoting a more sustainable society. In this context, materials science research has been focusing on the use of cellulose, the main constituent of plant cell walls, and more specifically cellulose nanocrystals (CNC).[1-3] CNCs consist of the crystalline part of cellulose and combine high surface area with enhanced mechanical properties, biodegradability and versatile surface chemistry. Studies have shown[4-5] in addition that the CNC surface is a unique platform for chemical or physical modification thanks to the presence of electrostatic charges and hydroxyl groups, allowing versatile CNC functionalization. These attributes make cellulose nanocrystals suitable for various advanced applications, such as reinforcement of composite materials, coatings, packaging, water filtration, biomedicine or electronic systems.[6-7]

A modification of the surface of cellulose, including that of CNCs by low molecular weight ligands[8-10] or polymer chains[11-13] is often required to adapt its properties to the desired applications. Polymers are bound to cellulose either by controlled surface polymerization[13-15] or by physical adsorption.[16] Physical adsorption remains an attractive route, as it is done in aqueous solvents under mild temperature and pressure conditions.[17] Beyond the use of ligands or polymers for coating cellulose, recent studies have focused on the association of CNCs with sub-micronic polymer particles, such as latexes.[18-19] One example of applications of latex/CNC assembly is the mechanical reinforcement of hydrophobic polymer matrices. In this context, latexes have been found to enhance the compatibility of cellulose nanofibrils or CNCs in thin films made from hydrophobic polymers, resulting in bio-composites of increased stiffness.[20-21] In addition, amphiphilic latexes have been shown to prevent the agglomeration of CNCs upon incorporation into the matrix and allow the production of composites with a higher CNC content.[22] This type of approaches has been used for developing nanopapers[23], packaging materials[24], and optical devices.[25] The ability of CNCs to form nematic and cholesteric liquid crystalline phases in aqueous media at high volume fraction has also received attention. Latex particles were found to improve the mechanical properties of composite films. With appropriate treatments, these films display remarkable fluorescence, birefringence and circular dichroism properties.[26-27] Additional applications of latex/CNC systems deal with the development of sag resistance waterborne acrylic latex,[28] the rheological control of pigments coating colors[29] or the stabilizers of Pickering emulsions.[30] Another application of CNCs, which we looked at recently, was to use CNCs as a cotton model, allowing studies to be done in bulk rather than at interfaces. In a collaborative work with Solvay, we defined various assessment protocols for bulk dispersions to test softener formulations made from double-tail cationic surfactants and biopolymers.[31-32] These studies revealed that a series of softeners formulations could be sorted out according to their affinity with cellulosic substrates, and that this affinity matched well with the softness performances.[31,33-34]

More generally, recent years have seen an upsurge in studies on the interaction between latex particles and cotton fibers, the objective being to make fabrics water-repellent, superhydrophobic, antibacterial or with specific finishing properties.[35-39] Along these lines, Carlson *et al.* reported the interaction of cationic latexes made by polymerization-induced self-assembly on cellulose model surface using quartz crystal microbalance and atomic force microscopy (AFM), as well as the formation of bio-composites films with adjustable hydrophobicity.[14,40] The adsorption experiments were successfully repeated using cellulose filter papers, of lower charge density and higher roughness. The influence of a polycation shell surrounding the latex particles was also investigated, revealing interparticle diffusion of polymer chains and strong adhesive joint.[41]

From these surveys, it appears that the potential applications of polymer particle/CNC composites, whether as solid materials or as colloids, requires systematic studies to understand their interfacial and phase behavior properties. The present work reports the study of cationic latex particles in the size range 200-300 nm synthesized using the RAFT/MADIX-mediated surfactant free emulsion polymerization method with cellulose nanocrystals. Dynamic light scattering, $\zeta$-potential, optical and electron microscopy revealed the structural properties of the pristine colloids, as well as of the latex/CNC hybrids. For latexes with low crosslinking density, it is noted that CNC adsorption induces a strong deformation of the spherical particles, which then take the form of faceted polyhedra. We also provide a new technique for labeling latex particles with the insertion of carbocyanine green dyes (PKH67) in the hydrophobic cores.[42-43] The PKH67 labeling is exploited to visualize and quantify the deposition of latexes on woven cotton fabrics. By varying the electrostatic charge on the latex particles, we found that the interaction with cellulose is strongest for latexes with the highest charge density. The present study demonstrates that CNCs can serve as model of micron sized fibers to assess the polymer deposition at cellulosic interfaces.

# 2. MATERIALS AND METHODS

## 2.1. Materials

Cellulose nanocrystals (CNC) were prepared according to earlier reports using catalytic and selective oxidation.[2-3] Briefly, cotton linters provided by Buckeye Technologies Inc. were hydrolyzed according to the method described by Revol *et al.* treating the cellulosic substrate with 65% (w/v) sulfuric acid at 63 °C during 30 min.[44] The suspensions were washed by repeated centrifugations, dialyzed against distilled water and sonicated for 4 min with a sonifier (Branson Ultrasonics Inc.) equipped with a 3 mm microtip. The suspensions were then filtered sequentially through a 8 and 1 µm cellulose nitrate membrane (Whatman). At the end of the process, a 2 wt. % aqueous stock dispersion (pH 4.5) was obtained, and later investigated using cryogenic transmission electron microscopy (cryo-TEM) and dynamic light scattering. Cryo-TEM reveals anisotropic particles in the form of laths which average length, width and thickness derived from image analysis were found equal to 180 ± 30 nm, 17 ± 4 nm and 7 ± 2 nm, respectively. Dynamic light scattering results displays a single relaxation mode in the autocorrelation function associated with a size distribution centered around 120 nm **(Supporting Information Fig. S1)**. Water was deionized with a Millipore Milli-Q Water system.

The polymer latexes used in this work are cationic core-shell particles kindly provided by Solvay. They were prepared through RAFT/MADIX-mediated surfactant free emulsion polymerization based on previously reported methodology.[40-41,45] RAFT/MADIX combines the reversible addition–fragmentation chain transfer (RAFT) polymerization and the macromolecular design by the interchange of xanthates (MADIX) to access well-controlled polymers and copolymers using polymerization-induced self-assembly (PISA) and emulsion polymerization.[46] PISA involves the chain extension in water of hydrophilic polymer chains terminated by a RAFT/MADIX agent (called macromolecular chain transfer agent of macro-CTA), the latter being synthesized by controlled radical polymerization beforehand with a hydrophobic monomer. After the pioneered work of Hawkett and coworkers,[47] the PISA methodology has been applied to a large range of macro-CTA and monomers.[19] The macromolecular chain transfer agents may also be introduced in water in combination with hydrophobic monomers to conduct surfactant-free emulsion polymerization. Here, the macro-CTAs were introduced in large amounts relative to the hydrophobic monomer, leading to particles with a hydrophobic dense core surrounded by a hydrophilic shell of macro-CTAs. In the present work, the hydrophilic block consisted of acrylamide (AM), a non-ionic monomer and a cationic monomer which is either (3-acrylamidopropyl)trimethylammonium chloride (APTAC) or N,N,N,N',N',N'',N''-heptamethyl-N''-3-(1-oxo-2-methyl-2-propenyl)aminopropyl-9-oxo-8-azo-decane-1,4,10-triammonium trichloride triquat (TQ). The latter monomer was synthesized by Solvay (reference WO 2003/101935). The hydrophobic monomers, butyl acrylate (BA) or 2-Ethylhexyl acrylate (2-EHA) are polymerized on the hydrophilic macro-chain transfer agent. During synthesis the polymers assemble in the form of spherical particles.

Three different polymers have been synthesized and assessed in the present study: P(AM-*co*-APTAC)-*b*-PBA, P(AM-*co*-APTAC)-*b*-PEHA and P(AM-*co*-TQ)-*b*-PBA. The polymers were provided in the form of a ~ 30 wt. % emulsion in water. The molecular weight, the fraction of cationic monomer in the macro-CTA $x_C$, and the weight ratio macro-CTA/hydrophobic block $R_{H/H}$ are provided in Table 1. Both APTAC-based copolymers have similar $X_C$ and $R_{H/H}$ values (10 and 20 respectively) whereas P(AM-*co*-TQ)-*b*-PBA is characterized by smaller values ($x_C$ = 3.3 and $R_{H/H}$ = 5). This later polymer is expected to exhibit less with CNCs or cotton fibers. The latex aqueous dispersions were purified by dialysis in Milli-Q water and were then diluted at 0.1 wt. %, and 0.005 wt. %. The pH of the dispersions was fixed at 6.

Woven cotton fabrics were used to investigate the latex particles deposition onto fibers. The fabrics are made of cotton yarns of diameters 250 μm, each of them being constituted of approximately 25 cotton fibers of average diameter 13 ± 3 μm (**Supporting information S2**). All fabrics were first treated with DI-water for 10 min and dried at 35 °C under air circulation before use. For the visualization of polymer deposits *via* fluorescence microscopy, a 5×10 mm² piece of cotton was dipped in a dispersion containing PKH67 labeled latexes at the concentration of 0.1 wt. % for 10 minutes. The fabric was then immersed for 60 s in water for rinsing and studied in wet and dried state.

**Table 1 :** *Polymer characteristics. $x_C$ denotes the fraction of cationic monomer in the macro-chain transfer agent $x_C$, and $R_{H/H}$ the weight ratio between the macro-CTA and the hydrophobic block.*

| Polymers | $x_C$ | $R_{H/H}$ |
|---|---|---|
| P(AM-*co*-APTAC)-*b*-PEHA | 10 | 20 |
| P(AM-*co*-APTAC)-*b*-PBA | 10 | 20 |
| P(AM-*co*-TQ)-*b*-PBA | 3.3 | 5 |

## 2.2 Mixing protocol

The interactions between latex particles and CNCs were investigated using the direct mixing formulation pathway. Latex and CNC batches were prepared in the same conditions of pH (pH 5) and concentration ($c$ = 0.005 wt. %) and then mixed at different ratios, noted $X = c_{Latex}/c_{CNC}$ where $c_{Latex}$ and $c_{CNC}$ are the latex and CNC weight concentrations in the mixtures, and the total concentration being $c = c_{Latex} + c_{CNC}$. The mixed dispersions were prepared in the range $X = 10^{-3}$ -$10^3$ at room temperature ($T$ = 25 °C). With this definition, the neat CNC (resp. latex) dispersion is characterized by X = 0 (resp. ∞). After mixing, the dispersions were stirred rapidly, let to equilibrate for 5 minutes and characterized by light scattering, $\zeta$-potential and optical microscopy.

## 2.3 Latex particles labelling and cotton treatment

Latex particles were labelled by using a fluorescent dye, PKH67 typically used in biology for cells staining. Its structure is given in **Supporting information S3**. For labelling 3 µl PKH67 stock solution was mixed rapidly with 100 µl of DI-water and vortexed for 10 s. It was then added to a 2700 µl of 0.1 wt. % latex dispersion. The final dye concentration was then $10^{-3}$ µM. Rapid vortexing for *ca.* 10 s was followed to ensure the dye insertion in the polymer particles. The dispersion was let to rest in the dark for 5 minutes.

## 2.4 Dynamic light scattering (DLS)

The scattering intensity $I_S$ and hydrodynamic diameter $D_H$ were measured using the Zetasizer NanoZS spectrometer (Malvern Instruments, Worcestershire, UK). A 4 mW He−Ne laser beam ($\lambda = 633$ nm) is used to illuminate the sample dispersion, and the scattered intensity is collected at a scattering angle of 173°. The second-order autocorrelation function $g^{(2)}(t)$ is analyzed using the cumulant and CONTIN algorithms to determine the average diffusion coefficient $D_C$ of the scatterers. The hydrodynamic diameter is then calculated according to the Stokes-Einstein relation, $D_H = k_B T / 3\pi\eta D_C$, where $k_B$ is the Boltzmann constant, $T$ the temperature and $\eta$ the solvent viscosity. Measurements were performed in triplicate at 25 °C after an equilibration time of 120 s. As CNCs are colloids with differing translational diffusion constants parallel and perpendicular to their main axis, the values obtained from DLS cannot be directly linked to the particle length, or to the hydrodynamic diameter. Following the suggestions from Ref.[4], the term hydrodynamic apparent particle size ($D_H^{app}$) will be used instead.

## 2.5. $\zeta$-potential

Laser Doppler velocimetry (Zetasizer, Malvern Instruments, Worcestershire, UK) using the phase analysis light scattering mode and a detection at an angle of 16 degrees was performed to determine the electrophoretic mobility and $\zeta$-potential of the different studied dispersions. Measurements were performed in triplicate at 25 °C, after 120 s of thermal equilibration.

## 2.6. Phase-contrast and fluorescence optical microscopy

Images were acquired on an IX73 inverted microscope (Olympus) equipped with a 60× objective. An Exi Blue camera (QImaging) and Metaview software (Universal Imaging Inc.) were used as the acquisition system. The illumination system "Illuminateur XCite Microscope" produced a white light, filtered for observing a green signal in fluorescence (excitation filter at 470 nm - bandwidth 40 nm and emission filter at 525 nm - bandwidth 50 nm). 8 microliters of the latex particle dispersion were deposited on a glass plate and sealed into a Gene Frame dual adhesive system (Abgene/Advanced Biotech). Images were digitized and treated by the ImageJ software and plugins (http://rsbweb.nih.gov/ij/).

## 2.7. Cryogenic transmission electron microscopy (Cryo-TEM)

Seven microliters of the samples were deposited on a lacey carbon coated 200 mesh (Ted Pella). The drop was blotted 3 times with a filter paper on a VitrobotTM (FEI). The grid was quenched rapidly in liquid ethane and cooled with liquid nitrogen to avoid the crystallization of the aqueous phase. The membrane was finally transferred into the vacuum column of a TEM microscope (JEOL 1400 operating at 120 kV) where it was maintained at liquid nitrogen temperature thanks to a cryo-holder (Gatan). The magnification was selected between 3000× and 40000×, and images were recorded with a 2k-2k Ultrascan camera (Gatan).

# 3. RESULTS AND DISCUSSION

### 3.1. Characterization of latex aqueous dispersions

The amphiphilic block copolymers, P(AM-*co*-APTAC)-*b*-PBA, P(AM-*co*-APTAC)-*b*-PEHA and P(AM-*co*-TQ)-*b*-PBA were synthesized by RAFT/MADIX-mediated surfactant free emulsion polymerization. These polymers were found to self-assemble during synthesis into sub-micrometer latex particles. Dilute aqueous dispersions were prepared and studied by dynamic light scattering, $\zeta$-potential and cryogenic transmission electron microscopy (cryo-TEM). DLS data show that the latex dispersions are stable over time (> months) and that the autocorrelation functions of the scattered light are associated with a single relaxation mode. This mode corresponds to hydrodynamic diameters $D_H$ around 200 - 300 nm and to dispersity indexes (pdi) of 0.05. **Figs. 1a-c** display the $D_H$ intensity distributions obtained from 0.1 wt. % dispersions for the three latexes, respectively. Additional scattering experiments were also performed as a function of the concentration and did not reveal aggregation. Sedimentation was not observed either. The excellent colloidal stability of the latex dispersions was attributed to the electrosteric effect provided by the positively charged shell around the particles.[48] Cryo-TEM reveals well-contrasted spherical particles, again uniform in size, as illustrated in **Figs. 1d-f**. The median diameters obtained from cryo-TEM were found in the range of 200 - 250 nm in good agreement with light scattering results. It is suggested that the difference in chain density between the core and shell chains account for the fact that the electronic contrast of the copolymer shell is significantly lower than that of PBA or PEHA cores, resulting in a lower capacity to scatter electrons.[40,49] Electrophoretic mobility measurements led to $\zeta$-potential values around +35mV, confirming that latexes are strongly charged and cationic. It is anticipated that the positive electrostatic charges in the shell will favor a strong attractive interaction with anionic CNCs and cotton fibers. Table 1 recapitulates the structural and electrostatic characteristics of the 3 latexes, as well as those of the cellulose nanocrystals for comparison. For the cellulose nanocrystals, the value of 120 nm obtained from DLS refers to the hydrodynamic apparent particle size $D_H^{app}$.[4]

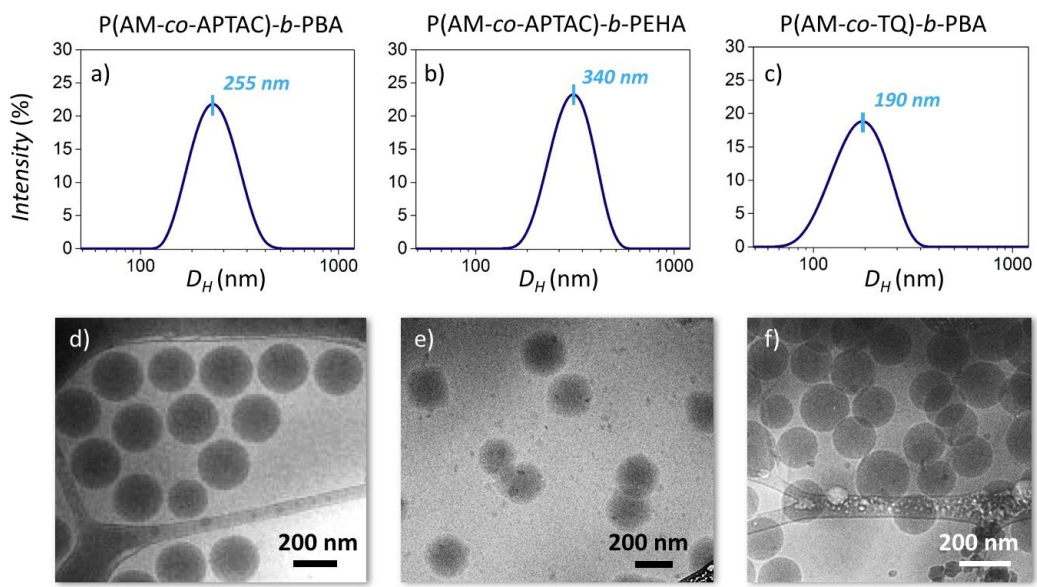

***Figure 1: a-c)*** *Hydrodynamic size distributions obtained from dynamic light scattering (DLS) and **d-f)** representative cryogenic electron microscopy (cryo-TEM) images of P(AM-co-APTAC)-b-PBA, P(AM-co-APTAC)-b-PEHA and P(AM-co-TQ)-b-PBA latex particles at 0.1 wt. %, respectively. The median diameters obtained from scattering and cryo-TEM are provided in Table 2.*

***Table 2:*** *List of the hydrodynamic and cryo-TEM diameters and $\zeta$-potential data for the three latex particles P(AM-co-APTAC)-b-PBA, P(AM-co-APTAC)-b-PEHA and P(AM-co-TQ)-b-PBA and for cellulose nanocrystals (CNCs).*

| Polymers | Dynamic light scattering | | Cryo-TEM | Electrophoresis |
|---|---|---|---|---|
| | $D_H$ (nm) | PdI | $D_{TEM}$ (nm) | $\zeta$ (mV) |
| P(AM-*co*-APTAC)-*b*-PBA | 255 | 0.04 | 220 | +38 |
| P(AM-*co*-APTAC)-*b*-PEHA | 340 | 0.05 | 250 | +40 |
| P(AM-*co*-TQ)-*b*-PBA | 190 | 0.06 | 200 | +28 |
| Cellulose nanocrystals (CNCs) | 120* | 0.15 | 200 | -29 |

*\*: Apparent hydrodynamic particle size*

### 3.2. Phase behavior of latex /cellulose nanocrystal aqueous dispersions

*3.2.1. P(AM-co-APTAC)-b-PBA latex/CNC induced aggregation*

In previous reports,[31-32] we used cellulose nanocrystals as a model of cotton to study cellulose/surfactant interactions. For this, we applied the continuous variation method developed by P. Job[50-51] and later modified by us for light scattering[52-54] to study the phase behavior of ternary dispersions.[31-32] Here, we implement this approach to assess latex/CNC interaction and co-assembly properties. **Fig. 2a** exhibits the hydrodynamic diameter $D_H$ obtained from mixtures of P(AM-*co*-APTAC)-*b*-PBA and CNC dispersions as a function of the concentration ratio $X = c_{Latex}/c_{CNC}$, where $c_{Latex}$ and $c_{CNC}$ denote the latex and CNC weight concentrations at a given $X$-value, respectively. The Job scattering plots are thereby constructed in a way that the total concentration $c = c_{Latex} + c_{CNC}$ (here 0.005 wt. %) is constant and remains in the dilute regime. In **Fig. 2a**, $D_H(X)$ is found to increase at low $X$, passes through a maximum at the critical value $X_C = 4 \pm 1$ and then decreases towards the diameter of latex particles. In Job scattering plots, such a feature is characteristic of strong interparticle interaction, and of the formation of mixed aggregates, here bewteen latexes and CNCs.[32,54] At the maximum, the $D_H$'s are larger than 1 µm and the aggregates precipitate. The precipitation range is pictured by the dashed area. The $\zeta$-potential data are given in **Fig. 2b**, which shows a progressive shift from negative ($\zeta = -29$ mV) for pristine CNCs to positive ($\zeta = + 38$ mV) for P(AM-*co*-APTAC)-*b*-PBA latexes. The point of zero charge is found at $X_{PZC} = 3 \pm 1$, in good agreement with the position of the light scattering peak. The results are interpreted in terms of electrostatic co-assembly associated with the process of charge titration and neutralization.[55] Apart from the stoichiometric, the aggregates are charged (negative for $X \ll X_C$ and positive for $X \gg X_C$) and repel each other, preventing further growth. Job scattering plots are valuable tools as they allow to identify concentration and mixing ranges where the dispersions are stable, a property that will prove crucial for cryo-TEM. A schematical representation of the structures obtained in each regime are depicted in the insets of **Fig. 2a**.

### 3.2.2. Imaging aggregates

In these experiments, the latex particles were labeled using the carbocyanine molecule PKH67 prior to the mixing with CNCs. This fluorescent dye emitting in the green ($\lambda$ = 525 nm) is commonly used in cellular biology for membrane imaging.[42] Here we applied protocols developped by us for surfactant and lipid vesicles.[56] As shown below, this protocol worked well with the latexes, revealing that the fluorophores spontaneously inserted in the hydrophobic latex cores. The micron-sized aggregates obtained at the critical $X_C$ are visualized by phase contrast and fluorescence optical microscopy (magnification ×20) in **Fig. 2c and 2d**, respectively. With fluorescence, an intense green signal is observed and found to co-localize with the aggregates imaged in phase contrast. Isolated particles were not observed in the many images collected, suggesting that the latexes have all interacted with the CNCs. **Fig. 2e** shows images of cuvettes at room temperature containing mixed dispersions in the range $X$ = 0.5 − 10. Turbid phases are observed for $X$ = 1 to 3, whereas at higher ratios precipitation is observed. The above results suggest strong attractive interactions between the oppositely charged latex particles and CNCs, even at concentrations as low as $c$ = 0.005 wt. %.

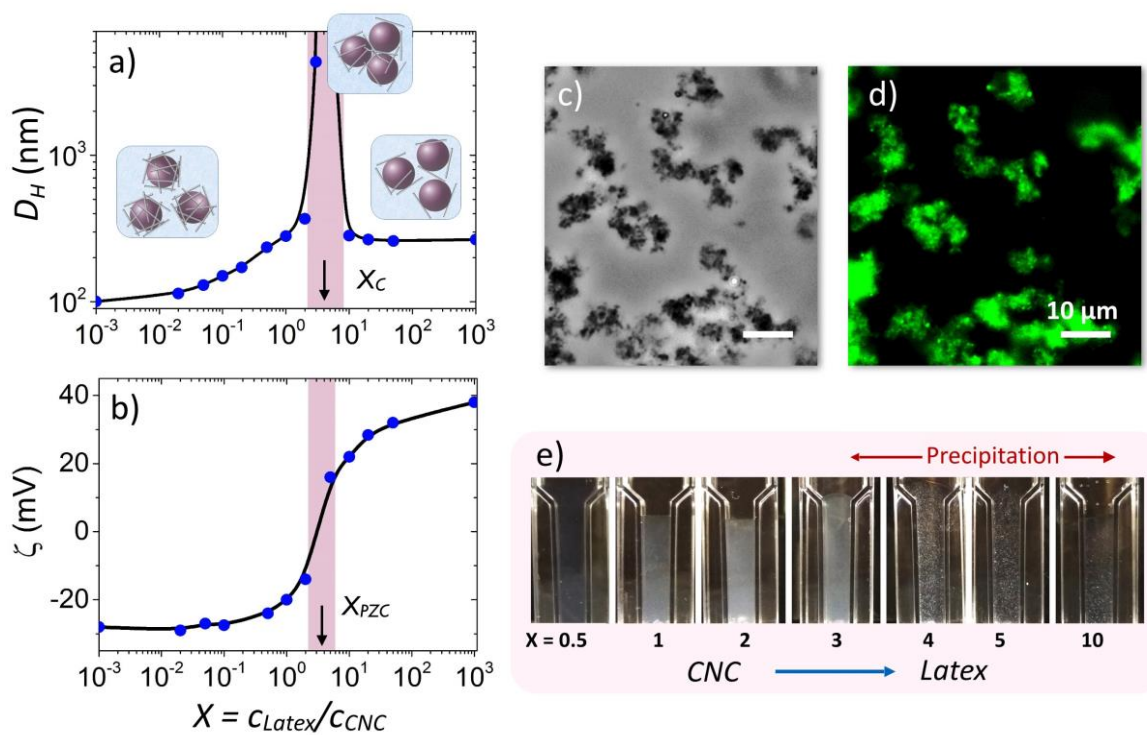

**Figure 2:** a) Hydrodynamic diameters and b) $\zeta$-potentials obtained for P(AM-co-APTAC)-b-PBA/CNC aqueous dispersions as a function of the mixing ratio $X = c_{Latex}/c_{CNC}$, where $c_{Latex}$ and $c_{CNC}$ denote the latex and CNC weight concentrations at a given $X$-value, respectively. c) Phase contrast and d) fluorescent microscopy images of mixed latex/CNC at $X$ = 3. e) Images of UV-visible cuvettes filled with dispersions in at different $X$'s between 0.5 and 10. The total concentration of the dispersions shown in these figures is 0.005 wt. %

### 3.2.3. Comparison between the different latex particles

Similar studies were conducted with the P(AM-*co*-APTAC)-*b*-PEHA and P(AM-*co*-TQ)-*b*-PBA latexes, again at the concentration of 0.005 wt.% and room temperature. As for P(AM-*co*-APTAC)-*b*-PBA, the hydrodynamic diameter exhibits a marked maximum with increasing $X$ (**Fig. 3a**), the position of which being correlated with the point of zero charge found by electrophoretic mobility measurements (*i.e.* $X_C \sim X_{PZC}$, **Fig. 3b**). Differences are that the $D_H$ maxima are less pronounced, and that their positions are shifted towards the higher $X$, that is towards higher particle content. More specifically, we found critical mixing ratios $X_C = 3$ for P(AM-*co*-APTAC)-*b*-PBA, $X_C = 7$ for P(AM-*co*-APTAC)-*b*-PEHA and $X_C = 40$ for P(AM-*co*-TQ)-*b*-PBA. The shift in the position of the maximum indicates that for these two latexes, more particles are required to reach the point of zero charge, and that these particles are less charged than the P(AM-*co*-APTAC)-*b*-PBA, typically by a factor 7/3 and 40/3 respectively. An analysis using polyelectrolyte assisted charge titration spectrometry has shown that the position of the DLS peak correlates with the structural charge density of the particles.[53] These outcomes suggest that both the nature of the hydrophobic block (PBA *versus* PEHA) and the charge density of the shell (APTAC *versus* TQ) determine the interaction with the cellulose nanocrystals.

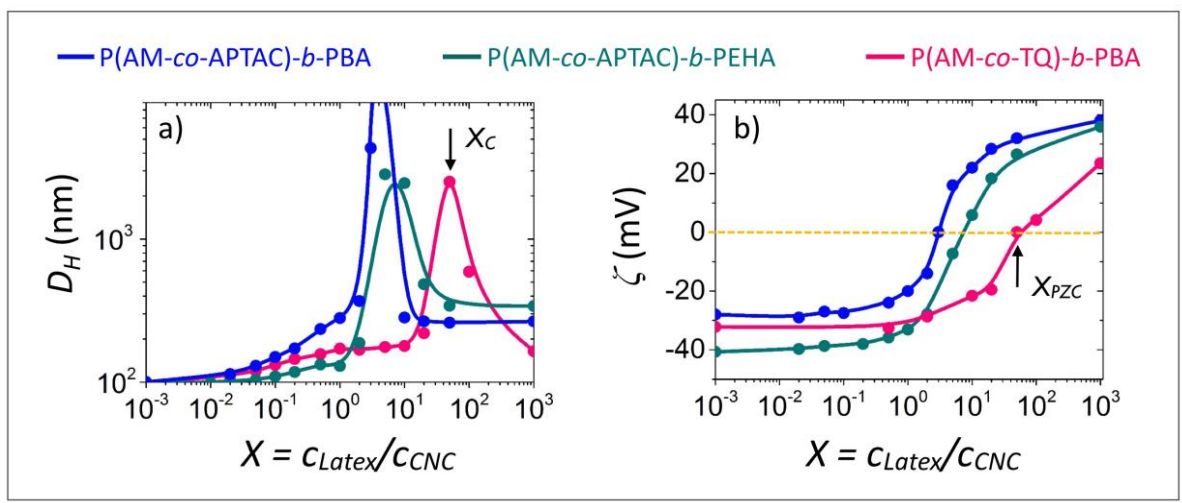

**Figure 3**: *a) Hydrodynamic diameter and b) $\zeta$-potential of mixed latex/CNC aqueous dispersions as a function of the mixing ratio $X = c_{Latex}/c_{CNC}$ for P(AM-co-APTAC)-b-PBA (blue symbols), P(AM-co-APTAC)-b-PEHA (green symbols) and P(AM-co-TQ)-b-PBA (red symbols). Each data point represents a different dispersion with a total concentration $c_{Latex}(X) + c_{CNC}(X)$ of 0.005 wt. %.*

### 3.2.4. Cryogenic transmission electron microscopy (cryo-TEM)

The cryo-TEM technique is of great utility in identifying the fine structure of colloids dispersed in a fluid. However, it has limitations, especially regarding the maximum size of objects that can be viewed. For non-deformable colloids, this size is estimated at 500 nm. It was hypothesized that colloids larger than 500 nm are expelled from the water film during blotting.[57] For this reason, the concentrations $c$ and mixing ratios $X$ of the latex/CNC dispersions had to be finely adjusted, in particular on the basis of the Job scattering results obtained for each polymer. For the P(AM-*co*-APTAC)-*b*-PBA and P(AM-*co*-APTAC)-*b*-PEHA, the total concentration $c$ was set at $c$ = 0.1 wt. % and the mixing ratios at $X$ = 0.2 and 20. For P(AM-*co*-TQ)-*b*-PBA, the high CNC sample was set at $X$ = 40 instead, the other parameters remaining unchanged. **Figs. 4a** display a series of cryo-TEM images for P(AM-*co*-APTAC)-*b*-PBA/CNC dispersions at $X$ = 0.2 (in **Fig. 4a1-a3**) and $X$ = 20 (**Fig. 4a4-a5**). In the conditions used, the majority of the CNCs are found to be adsorbed at the latex surfaces, building an entangled array of nanofibers around.

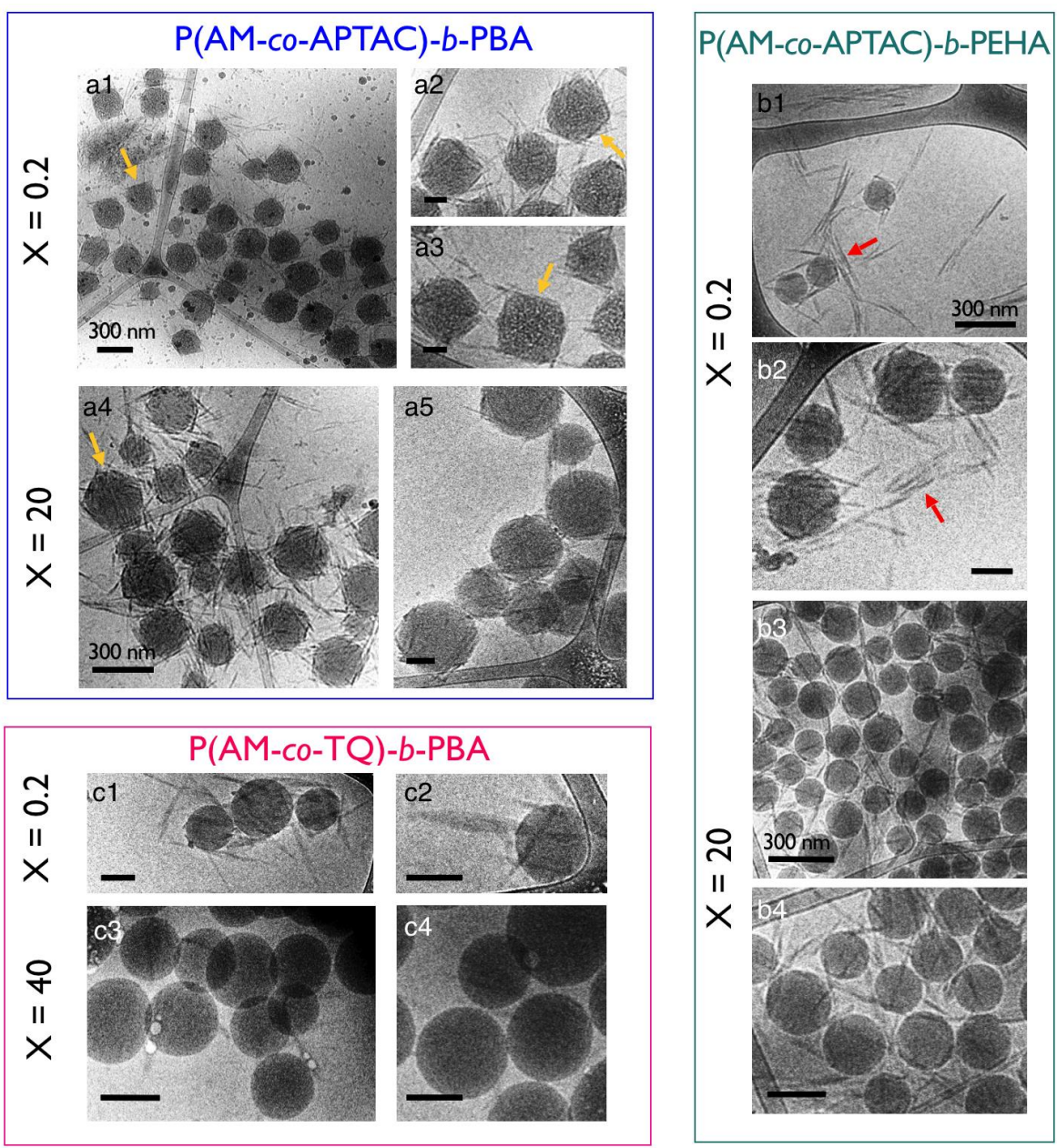

**Figure 4**: *Cryo-TEM images of latex/CNC dispersions at the concentration of 0.1 wt. % with a) with P(AM-co-APTAC)-b-PBA, b) P(AM-co-APTAC)-b-PEHA and c) P(AM-co-TQ)-b-PBA. The scale bars are 100 nm unless otherwise specified.*

It is assumed that this CNC array, connecting different particles is responsible for the latex aggregation observed in light scattering and microscopy. For both $X$'s, free CNCs could not be found in the many images collected. A close inspection of the particle surface reveals approximately 10 nanocrystals per object for $X$ = 0.2 and 20. This feature fits well with the fact that in electrostatic complexation schemes, co-assembly occurs at a given stochiometric charge ratio, as has already been demonstrated for nanoparticles and flexible polymers.[53,58-59] The cryo-TEM images also show that latexes with many CNCs around are deformed, taking the shape of faceted polyhedrons, such as cubes (**Fig. 4a3**) or pieces of octahedron (**Fig. 4a4**), as indicated by arrows. This deformation is assumed to result from the competition between the elastic deformation energy of the particles and the adhesion with the rigid and thread-like nanofibers. Experiments conducted with P(AM-*co*-APTAC)-*b*-PEHA provides similar characteristics to those of P(AM-*co*-APTAC)-*b*-PBA. For this polymer, a strong adsorption of the CNCs at the particle surface is observed (**Fig. 4b1-4b4**). Differences concern the average number of nanocrystals per particle, of the order of 5 and the fact that the latex spheres are not as deformed as in the previous case. This later observation is consistent with the greater rigidity of the PHEA cores compared to APTAC cores. In the case of P(AM-*co*-TQ)-*b*-PBA (**Figs. 4c**), CNCs still adsorb to the latex surface, but in fewer numbers than in previous cases. Here again, the particles do not present any noticeable deformation. In conclusion, for strongly charged surfaces and soft latex cores, *i.e.* for P(AM-*co*-APTAC)-*b*-PBA, a large number of CNCs adhere to the particles and induce a deformation of the initially spherical shape. For more rigid cores, or for weaker interactions, the particles do not deform. These differences could be attributed to the higher crosslinking degree expected for PHEA due to branching and hydrogen bonds.

### 3.3. Adsorption of latex particles on woven cotton fabrics

To assess latex interactions at cellulosic interfaces, particle adsorption was monitored on woven cotton fabrics made from microfibers. Studies have shown that the $\zeta$-potential of cellulose microfibers is lower than that of the CNCs, around -5 mV for the first and -30 mV for the second.[60] It is therefore important to evaluate to what extent a lower charge density may affect the latex deposition on cotton. The woven structure put under scrutiny here is illustrated in **Fig. 5a** using scanning electron microscopy. This piece of fabric is characterized by different length scales, which are the average fiber diameter, the size of the voids between yarns and the yarn diameter (**Fig. 5b**). The characteristic sizes are respectively around 10, 100 and 250 µm.

In a typical experiment, a 30 mg piece of cotton of size 5×10 mm$^2$ is dipped in 3 mL of a dispersion containing PKH67 labeled latexes at the concentration of 0.1 wt. % for 10 minutes. The fabric is then immersed for 60 s in water for rinsing and let dried at 35 °C for one hour. **Figs. 5c-e** display phase-contrast (upper panel) and fluorescent (lower panel) microcopy views of fabric pieces treated with P(AM-*co*-APTAC)-*b*-PBA, P(AM-*co*-APTAC)-*b*-PEHA and P(AM-*co*-TQ)-*b*-PBA latexes, respectively. The fluorescence images are taken after drying at exposure times of 100, 200 and 1000 ms and they are shown without contrast adjustment. The 200 µm sized yarns are recognizable and separated by large voids, in agreement with the SEM results. In the 525 nm emission mode, the fibers exhibit a fluorescence signal, illustrating that PKH67 labeled latexes have been adsorbed on cotton in a significant manner. Close-up views of the fibers taken with an ×40 objective show that the deposition is more uniform in the dried state compared to that of the wet state (**Supporting information S4**). Control experiments with unlabeled latexes and with pristine PKH67 dispersions show that in these conditions the fabrics are not fluorescent.[56] **Figs.**

**5c-e** also reveal that the fluorescence signal decreases noticeably from P(AM-*co*-APTAC)-*b*-PBA to P(AM-*co*-APTAC)-*b*-PEHA and then to P(AM-*co*-TQ)-*b*-PBA. Similar outcomes were obtained in the wet state (**Supporting information S5**). These results fit well with the overall picture obtained from the interaction studies between CNCs and latexes, namely that the adsorption of latex particles decreases with the cationic charge density, and that the interaction is mainly driven by electrostatics.

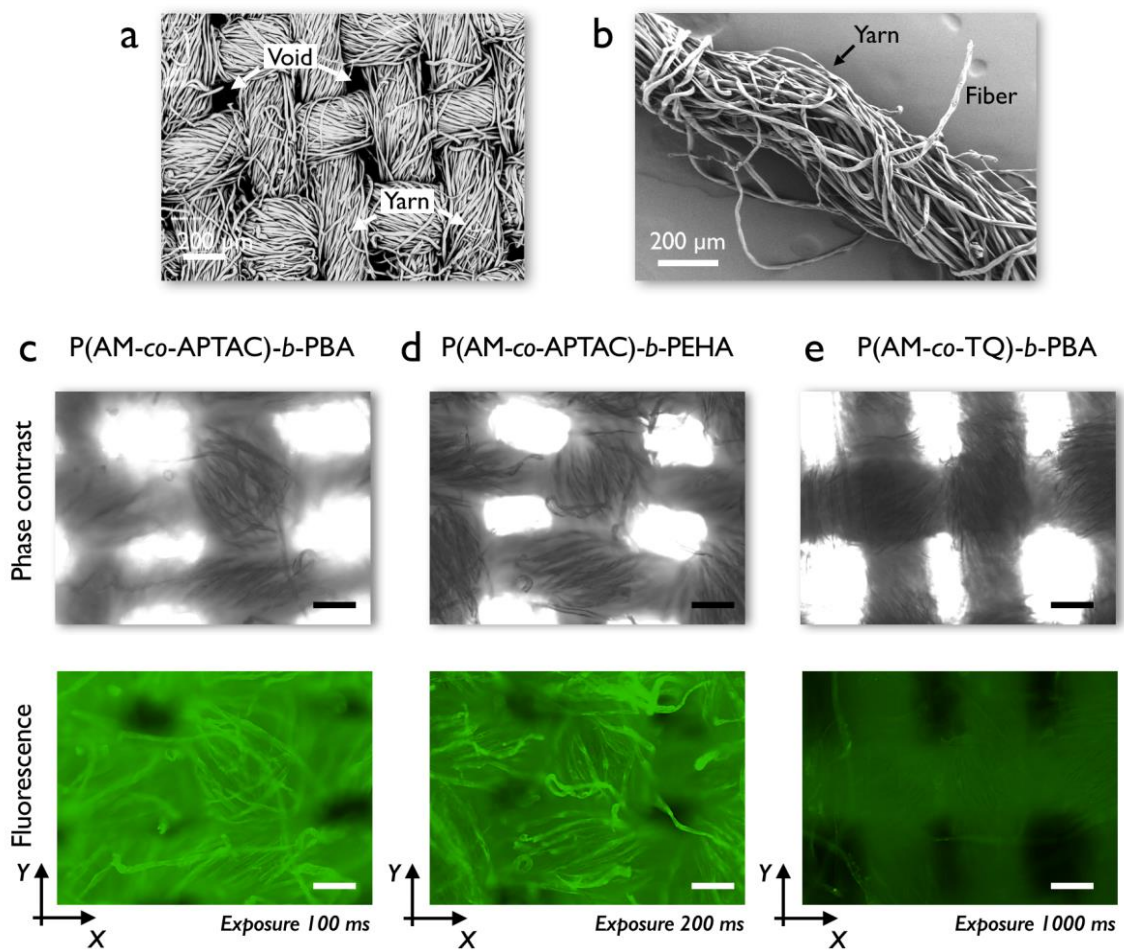

**Figure 5: a)** *Scanning electron microscopy image of a piece of cotton fabric used in this work.* **b)** *Scanning electron microscopy (SEM) image of an individual yarn.* **c), d)** *and* **e)** *Pieces of cotton fabric treated with fluorescent latex dispersions using P(AM-co-APTAC)-b-PBA, P(AM-co-APTAC)-b-PEHA and P(AM-co-TQ)-b-PBA, respectively and observed by optical microscopy. The upper panels show phase contrast images, whereas the lower panels show the fluorescence images of the same locations. Observations were made in the dried state at magnification ×10 (scale 100 μm).*

To determine the mass of polymers deposited on cotton, the spatial distribution of the fluorescent intensity was quantified, allowing a comparison of the affinity of latex particles to the cellulosic surface. To this aim, fluorescence data of a piece of fabric was acquired at increasing exposure times between 10 and 1000 ms, revealing a linearity of the green signal with the time (**Supporting information S6**). This led to a fluorescence intensity expressed in counts per millisecond and per pixel, which depends primarily on the adsorbed amount. **Fig. 6a** displays the integrated fluorescence intensity integrated along the Y-axis and as a function of the distance along the X-axis, both directions being defined in **Fig. 5**. The data emphasizes a constant signal as a function of the distance along the piece of tissue. Its small spatial fluctuations around an average value indicate a uniform latex deposition on microfibers.

For the conversion of the fluorescence intensity in mass per unit area $m_A$, a calibration experiment was performed to evaluate the fluorescent contrast of each labeled particle. A 5 µL drop of the latex dispersions at 0.1 wt. % was deposited on a glass substrate and let dried in an oven for one hour. $m_A$-values on silica was estimated at $0.035 \pm 0.005$ mg cm$^{-2}$ for the three polymers. Treated in the same way as the cotton fabrics, *i.e.* collecting fluorescence data at increasing exposure times, the coefficient linking fluorescence intensity (in cts pixel$^{-1}$ ms$^{-1}$) and the mass of polymers per unit (in mg cm$^{-2}$) area could be retrieved (**<span style="color:red">Supporting information S7</span>**). These coefficients were found to be close to each other, at 2.87, 2.40 and 3.85 for P(AM-*co*-APTAC)-*b*-PBA, P(AM-*co*-APTAC)-*b*-PEHA and P(AM-*co*-TQ)-*b*-PBA latexes, respectively, indicating that the coupling of carbocyanine PKH67 dyes to the particles was only slightly dependent on the polymers. In **Fig. 6b**, we found $m_A$-values in the range 0.5-5 mg cm$^{-2}$. Of note, these values are much higher than the detection limit of the method described previously, estimated at 10$^{-3}$ mg cm$^{-2}$. For P(AM-*co*-APTAC)-*b*-PBA, $m_A$ is 3 and 6 times larger than those of P(AM-*co*-APTAC)-*b*-PEHA and P(AM-*co*-TQ)-*b*-PBA, respectively. These results confirm that the highest adsorption on cotton is found for the latex particle with the highest electrostatic charge, and show that our fluorescence technique allows the visualization and quantification of adsorbed deposits.

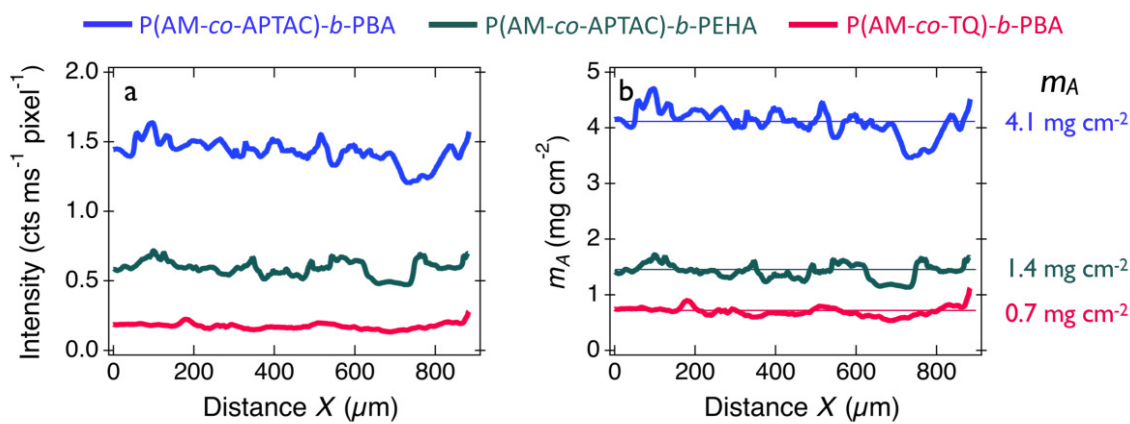

**Figure 6: a)** *Fluorescence intensity obtained from woven cotton fabrics immersed in P(AM-co-APTAC)-b-PBA, P(AM-co-APTAC)-b-PEHA and P(AM-co-TQ)-b-PBA latex dispersions and let dried for one hour in controlled environment. The intensity is expressed in counts per millisecond of exposure time and per pixel. The curves correspond to the images shown in Figs. 5c, 5d and 5e, lower panels. The curves display the intensity along the X-direction and integrated over the Y-direction, as explained in the text.* **b)** *Mass of polymers deposited on cotton fabrics per unit area calculated form the data in Fig. 6a, showing that the highest deposition is obtained for the most cationic latex particles, P(AM-co-APTAC)-b-PBA.*

# 4. CONCLUSION

In this work we evaluate the interactions of latex particles synthesized using the RAFT/MADIX-mediated surfactant free emulsion polymerization with cellulosic interfaces. These interfaces are those of cellulose nanocrystals (CNCs) and of microfibers of woven fabrics. Compared to nanocrystals, cotton microfibers have 1000 larger diameters as well as lower $\zeta$-potential values. Latex particles with poly(butyl acrylate) or poly(2-EHA) hydrophobic cores of different cross linking densities and different electrostatic charge densities in the shell were prepared. The latexes were first assessed in solution with CNCs following protocols capable of evaluating the interaction strengths between the oppositely charged CNCs and latexes. We show that the latex/CNC assembly is mediated by electrostatic interaction, this interaction being the strongest with the most charged particles, namely with P(AM-$co$-APTAC)-$b$-PBA. The main result that emerges from this study is the observation in cryo-TEM of latex particles coated with entangled arrays of CNCs, and for the softest particles of a notable deformation of their structure. Cryo-TEM shows that latexes with many CNCs around them are deformed, taking the shape of faceted polyhedrons. The deformation is assumed to result from the competition between the elastic energy of the soft particles and the interaction with the rigid and thread-like nanofibers. The second important outcome deals with the adsorption properties of the latex particles on cotton tissues. Using a fluorescence labeling technique, the distribution of the particles adsorbed on millimeter-sized surfaces of fabric could be visualized and quantified. It is found that in the conditions used, the spatial distribution of deposited latexes is uniform in the dried state, and that the mass of deposited materials per unit area reaches values around 1 mg cm$^{-2}$. It should be added that the fluorescence technique proposed here is quite sensitive, as it is able to detect values down to $10^{-3}$ mg cm$^{-2}$. We also show a good correlation between the bulk and the surface affinity of latexes for cellulose interfaces. In the two sets of experiments conducted, we found that the latex-cellulose interaction is strongest for latexes with the highest charge density, here P(AM-$co$-APTAC)-$b$-PBA. The present study demonstrates that CNCs can serve as model of micron sized fibers to assess the deposition of polymers at cellulose interfaces. It also constitutes a working guide for tailoring new latex/CNC structures, opening avenues in the synthesis of composite materials, or in fiber coating.

# ASSOCIATED CONTENT

## Supporting information

The Supporting Information is available free of charge on the ACS Publications website at DOI: 10.1021/acsapm.xxxxx.

S1: Cryo-TEM and dynamic light scattering characterization of cellulose nanocrystals; S2: Diameter distribution of the cotton microfibers used in this study; S3: – Molecular structure of the PKH67 carbocyanine dye used for labelling the latex particles; S4 – Fluorescent images of cotton microfibers after treatment with PKH67 labeled latexes using X40 objective; S5 – Fluorescence intensities obtained for woven cotton fabrics impregnated with PKH67 labeled latexes in the wet state; S6 – Quantification of the fluorescence signal using increasing illumination times in fluorescent optical microscopy; S7 – Quantifying the mass of polymers deposited on a substrate from its fluorescence intensity

# AUTHOR INFORMATION

## Corresponding Author

E-mail: jean-francois.berret@u-paris.fr,  Tel.: +33 1 5727 6147
Université de Paris, CNRS, Matière et systèmes complexes, 75013 Paris, France
ORCID Jean-François Berret: 0000-0001-5458-8653

## Authors

**Evdokia K. Oikonomou**
Université de Paris, CNRS, Matière et systèmes complexes, 75013 Paris, France, ORCID Evdokia K. Oikonomou: 0000-0003-2690-0982
**Konstantin Golemanov**
Solvay Research & Innovation Center Singapore, 1 Biopolis Drive, Amnios, Singapore 138622
**Pierre-Emmanuel Dufils**
Solvay Paris, 52 rue de la Haie-Coq, 93308 Aubervilliers Cedex, France
**James Wilson**
Solvay Paris, 52 rue de la Haie-Coq, 93308 Aubervilliers Cedex, France
**Ritu Ahuja**
Solvay Research & Innovation Center Singapore, 1 Biopolis Drive, Amnios, Singapore 138622
**Laurent Heux**
Centre de recherches sur les macromolécules végétales (CERMAV), BP 53, 38041, Grenoble, France, ORCID Laurent Heux: 0000-0002-1934-4263

## Author Contributions

E.K.O., K.G., P.-E. D. and J.-F. B. conceived the research. E.K.O. prepared the samples and performed all experiments on latexes, CNCs and fabrics. E.K.O. and J.-F. B. wrote the manuscript. All authors have given approval to the final version of the manuscript.

## Notes

The authors declare no competing financial interest.

# ACKNOWLEDGMENTS

We thank Annie Vacher and Marc Airiau from Solvay, Auberviliers for the cryo-TEM experiments and for the fruitful discussions. Olivier Onnaïnty and Thibaud Carouhy (Solvay) are also recognized for their contributions to the synthesis of polymers. Sarra Gam Derouich from the Laboratoire ITODYS (Université de Paris) is acknowledged for the scanning electron microscopy experiments on cotton fabrics. ANR (Agence Nationale de la Recherche) and CGI (Commissariat à l'Investissement d'Avenir) are gratefully acknowledged for their financial support of this work through Labex SEAM (Science and Engineering for Advanced Materials and devices) ANR 11 LABX 086, ANR 11 IDEX 05 02. This research was supported in part by the Agence Nationale de la Recherche under the contract ANR-13-BS08-0015 (PANORAMA), ANR-12-CHEX-0011 (PULMONANO), ANR-15-CE18-0024-01 (ICONS), ANR-17-CE09-0017 (AlveolusMimics) and by Solvay.

# ABBREVIATIONS

2-EHA, 2-Ethylhexyl acrylate
AM, acrylamide
APTAC, (3-acrylamidopropyl)trimethylammonium chloride; TQ, N,N,N,N',N',N'',N''-heptame BA, butyl acrylate
CNC, cellulose nanocrystals
Cryo-TEM, cryogenic transmission electron microscopy
DLS, dynamic light scattering
thyl-N''-3-(1-oxo-2-methyl-2-propenyl)aminopropyl-9-oxo-8-azo-decane-1,4,10-triammonium trichloride triquat
macro-CTA, macro-chain transfer agent
MADIX, macromolecular design by interchange of xanthate.
PISA: polymerization-induced self-assembly
RAFT, reversible addition–fragmentation chain-transfer

## TOC IMAGE

## Electrostatic-driven adsorption of cellulose nanocrystals on polymer latexes

Cellulose nanocrystals

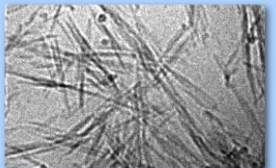

Polymer latex

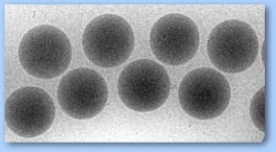

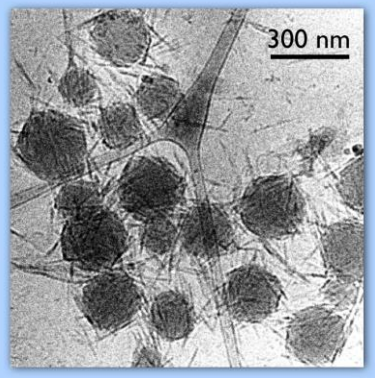